\documentclass[twocolumn,showkeys,prl,amsmath,amssymb]{revtex4}

\usepackage{graphicx}
\usepackage{dcolumn}
\usepackage{bm}

\newcommand{\simle}
{\raisebox{-0.75ex}[-1.5ex]{$\;\stackrel{<}{\sim}\;$}}

\def\e{{\epsilon}}
\def\k{{ {\bf k} }}

\def\w{{\omega}}

\def\g{{\gamma}}

\begin{document}


\title{Theory of Thermal Conductivity in High-$T_{\rm c}$ Superconductors
below $T_{\rm c}$: \\
Comparison between Hole-Doped and Electron-Doped Systems}

\author{Hideyuki {\sc Hara} and Hiroshi {\sc Kontani}}

\address{Department of Physics, Nagoya University,
Furo-cho, Nagoya 464-8602, Japan.}
 
\date{\today}

\begin{abstract}
In hole-doped high-$T_{\rm c}$ superconductors, 
thermal conductivity $\kappa$ increases drastically just below 
$T_{\rm c}$, which has been considered as a hallmark of a nodal gap.
In contrast, such a coherence peak in $\kappa$ is not visible
in electron-doped compounds, which may indicate a full-gap state 
such as a $d+is$-wave state.
To settle this problem, we study $\kappa$ in the Hubbard model
using the fluctuation-exchange (FLEX) approximation, which 
predicts that the nodal $d$-wave state is realized 
in both hole-doped and electron-doped compounds.
The contrasting behavior of $\kappa$ in both compounds 
originates from the differences in the hot/cold spot structure.
In general, a prominent coherence peak in $\kappa$ appears
in line-node superconductors {\it only when the cold spot exists 
on the nodal line.}
\end{abstract}

\sloppy

\maketitle


In strongly correlated electron systems, transport phenomena give us 
significant information on the many-body electronic states.
In high-$T_{\rm c}$ superconductors (HTSCs), for example,
both the Hall coefficient $R_{\rm H}$ and the thermoelectric power $S$
are positive in hole-doped compounds such as YBa$_2$Cu$_3$O$_{7-\delta}$ 
(YBCO) and La$_{2-\delta}$Sr$_\delta$CuO$_4$ (LSCO),
whereas they are negative in electron-doped compounds like
Nd$_{2-\delta}$Ce$_\delta$CuO$_4$ (NCCO) and 
Pr$_{2-\delta}$Ce$_\delta$CuO$_4$ (PCCO)
 \cite{Sato}.
These experimental facts originate from the 
difference in the ``cold-spot,'' which is the portion of the 
Fermi surface where the relaxation time of a quasiparticle (QP),
$\tau_\k$, takes the maximum value \cite{Kontani-Hall,Kontani-S,Kontani-rev}:

Below $T_{\rm c}$, electronic thermal conductivity $\kappa$ 
has been observed intensively since it gives us considerable 
information on the superconducting state; it is the only 
transport coefficient which remains finite below $T_{\rm c}$.
For example, the $\k$-dependence of the SC gap can be determined by 
the angle resolved measurement of $\kappa$ under the magnetic field 
\cite{izawa,Matsuda-rev}.
Also, one can detect the type of nodal gap structure (full-gap,
line-node, or point-node) by measuring $\kappa$
at low temperatures ($T\ll T_{\rm c}$).
For $T\simle T_{\rm c}$, $\kappa$ also shows rich variety
of behavior in various superconductors.
In conventional full-gap $s$-wave superconductors, 
the opening of the SC gap rapidly decreases the density of 
thermally excited QPs, causing $\kappa$ to decrease.
On the other hand, $\kappa$ shows ``coherence peak'' behavior
just below $T_{\rm c}$ in several unconventional superconductors 
with line-node gaps, e.g., hole-doped HTSC \cite{Popoviciu,Ong,Ong2},
CeCoIn$_5$ \cite{Movshovich,Kasahara}, and URu$_2$Si$_2$ \cite{Matsuda}.
A previous theoretical study based on a BCS model with $d$-wave
pairing interaction \cite{Hirshfeld} discussed that 
the coherence peak in YBCO originates from the steep reduction in
$\tau$ below $T_{\rm c}$.

In sharp contrast, no coherence peak in $\kappa$ is observed in 
electron-doped HTSCs \cite{Cohn,Fujishiro}, 
irrespective that a recent ARPES measurement \cite{Takahashi} 
suggests that the $d_{x^2\mbox{-}y^2}$-wave state is realized.
The observed $\k$-dependence of the SC gap function in NCCO,
which prominently deviates from $\cos k_x - \cos k_y$, is well 
reproduced by the fluctuation-exchange (FLEX) approximation 
\cite{Hirashima}, which is a self-consistent spin fluctuation theory.
On the other hand, recent point-contact spectroscopy for PCCO \cite{Qazilbash}
suggests that a full-gap SC state such as $d_{x^2\mbox{-}y^2}+is$ or 
$d_{x^2\mbox{-}y^2}+id_{xy}$ state is realized for $\delta=0.15$ and 0.17.
To find out the real SC state in electron-doped HTSC,
we have to elucidate whether the ``absence of coherence peak in $\kappa$''
is a crucial hallmark of the full-gap SC state, or it can occur
even in nodal gap superconductors.

In this letter, we present a theoretical study of the electronic 
thermal conductivity $\kappa$ in HTSCs using the FLEX approximation.
This is the first numerical study of transport properties
in the SC state based on the repulsive Hubbard model.
In deriving the relaxation time $\tau_\k$, both the strong
inelastic scattering due to Coulomb interaction and weak
elastic impurity scattering are taken into consideration,
which corresponds to optimally-doped YBCO and NCCO samples, respectively.
We find that a sizable coherence peak of $\kappa$ in YBCO
originates from the reduction in inelastic scattering.
In contrast, the coherence peak is absent in NCCO in spite of that
$d_{x^2\mbox{-}y^2}$-wave SC state is realized,
since the nodal point does not coincide with the cold spot 
in the normal state.
Thus, contrasting behaviors of $\kappa$ in YBCO and NCCO are explained
on the same footing as $d_{x^2\mbox{-}y^2}$-wave superconductors. 
This result was not derived in the BCS model \cite{Hirshfeld}.

Here, we study the following repulsive Hubbard model:
\begin{eqnarray}
{\cal H}=\sum_{\mathbf{k},\sigma}\epsilon_{\mathbf{k}}
c^\dagger_{\mathbf{k}\sigma}c_{\mathbf{k}\sigma}+\frac{{\rm U}}{N}
\sum_{\mathbf{k,k'q}}c^\dagger_{\mathbf{k+q}\uparrow}
c^\dagger_{\mathbf{k'-q}\downarrow}c_{\mathbf{k'}\downarrow}
c_{\mathbf{k}\uparrow}
 \label{eqn:Ham}
\end{eqnarray}
where $U$ is the Coulomb interaction and $\epsilon_{\mathbf{k}}=2t_0
(\cos(k_x)+\cos(k_y))+4t_1\cos(k_x)\cos(k_y)+2t_2(\cos(2k_x)+\cos(2k_y))$
is the kinetic energy of free electrons.
Hereafter, we put $t_0=-1.0$, $t_1=0.167$, and $t_2=-0.2$ \cite{Kontani-Hall}
to reproduce the Fermi surface of YBCO and NCCO.
We also put ${\rm U}= 8.0$ for YBCO and ${\rm U}= 5.4$ for NCCO.
Here, no phenomenological fitting parameters
are introduced except for $U$.

\begin{figure}
\includegraphics[scale=0.33]{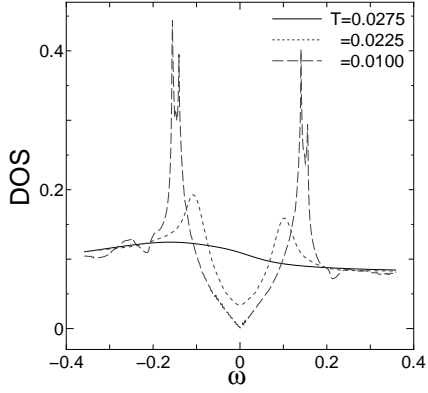}
\caption{Temperature dependence of the DOS
given by the FLEX approximation for YBCO ($n=0.9$).
Note that $T_{\rm c}=0.024$.}
\label{fig:DOS}
\end{figure}

In the FLEX approximation, the normal and anomalous self-energies
are given by
\begin{eqnarray}
\Sigma^n(\mathbf{k},i\e_n)&=&\displaystyle\frac{{\rm U}^2T}{N}
\sum_{\mathbf{k},l}G_{\mathbf{q+k}}(i\e_n+i\omega_l) \nonumber \\
& &\times\displaystyle\left(\frac{3}{2}\chi_s+\frac{1}{2}\chi_c-\chi_0
\right)_{\mathbf{q},\omega_l}
\label{eqn:Self-n} \\
\Sigma^a(\mathbf{k},i\e_n)&=&\displaystyle\frac{-{\rm U}^2T}{N}
\sum_{\mathbf{k},l}F^{\dagger}_{\mathbf{q+k}}(i\e_n+i\omega_l) \nonumber \\
& &\times\displaystyle\left(\frac{3}{2}\chi_s-\frac{1}{2}\chi_c-\phi_0
\right)_{\mathbf{q},\omega_l}
\label{eqn:Self-a}
\end{eqnarray}
where $\e_n= \pi T (2n+1)$ and $\w_l=2\pi T l$ are 
the Matsubara frequencies for fermions and bosons, respectively.
$\chi_s$ and $\chi_c$ are the dynamical spin and charge 
susceptibilities, which are given by
\begin{eqnarray}
& &\chi_s(\mathbf{q},i\omega_l)=\displaystyle\frac{\chi_0+\phi_0}{1-{\rm U}
(\chi_0+\phi_0)}
\label{eqn:chi-s} \\
& &\chi_c(\mathbf{q},i\omega_l)=\displaystyle\frac{\chi_0-\phi_0}{1-{\rm U}
(\chi_0-\phi_0)}
\label{eqn:chi-c} \\
& &\chi_0(\mathbf{q},i\omega_l)=-\displaystyle\frac{T}{N}\sum_{\mathbf{k},n}
G_{\mathbf{k+q}}(i\e_n+i\omega_l)G_{\mathbf{k}}(i\e_n)
\label{eqn:chi0} \\
& &\phi_0(\mathbf{q},i\omega_l)=-\displaystyle\frac{T}{N}\sum_{\mathbf{k},n}
F^{\dagger}_{\mathbf{k+q}}(i\e_n+i\omega_l)F_{\mathbf{k}}(i\e_n)
\label{eqn:phi0}
\end{eqnarray}
where $G$ and $F$ are the normal and anomalous Green function, respectively.
They are given by
\begin{eqnarray}
G(i\e_n)&=&(i\e_n+{\tilde \e}_\mathbf{k}+\Sigma^n(-i\e_n))D(\e_n)^{-1}
\label{eqn:GreenG} \\
F(i\e_n)&=&\Sigma^a(-i\e_n)D(i\e_n)^{-1}
\label{eqn:GreenF} \\
D(i\e_n)&=&(-\e_n-{\tilde \e}_\mathbf{k}+\Sigma^n(-i\e_n))
(-\e_n+{\tilde \e}_\mathbf{k}+\Sigma^n(i\e_n)) \nonumber \\
& &-(\Sigma^a(-i\e_n))^2
\label{eqn:Green}
\end{eqnarray}
where ${\tilde \e}_\mathbf{k}=\e_\mathbf{k}-\mu$; $
\mu$ is the chemical potential.
In the FLEX approximation, we solve eqs. (\ref{eqn:Self-n})-(\ref{eqn:Green})
self-consistently by choosing $\mu$ to adjust the electron filling $n$.
In the following numerical study, 
we use $64\times64$ $\k$-meshes and 2048 Matsubara frequencies.
Figure \ref{fig:DOS} represents the density of states (DOS);
$\rho(\w)= \frac1N \sum_\k G_\k^A(\w)/\pi$.
Here, the advanced (retarded) Green function $G_\k^A(\w)$ ($G_\k^R(\w)$)
is given by the numerical analytic continuation of the Matsubara Green function
from the lower (upper) half plane in the complex $\w$ space.

\begin{figure}
\includegraphics[scale=0.35]{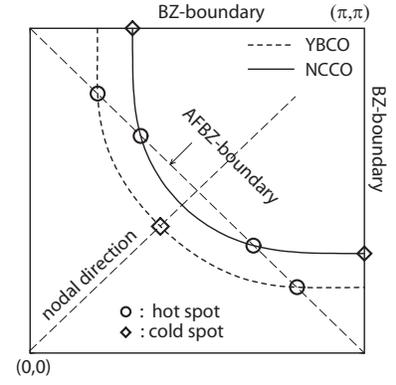}
\caption{Position of the hot/cold spots for YBCO ($n=0.9$) 
and for NCCO ($n=1.10$).
AFBZ represents the antiferromagnetic Brillouin zone.
}
\label{fig:FS} 
\end{figure}
Figure \ref{fig:FS} shows the location of the hot/cold spots
for both YBCO and NCCO in the normal state.
The transport phenomena are governed by QPs around the cold spot,
where the QP damping rate
$\gamma_\k={\rm Im}\Sigma_\k^n(-i\delta)$ [$=1/2\tau_\k$]
takes the minimum value.
According to the FLEX approximation,
the cold-spot in hole-doped [electron-doped] systems is around 
$(\pi/2,\pi/2)$ [$(\pi,0)$] \cite{Kontani-Hall}.
The position of the cold-spot in electron-doped systems
was confirmed by ARPES measurements \cite{Armitage1,Armitage2}
after the theoretical prediction \cite{Kontani-Hall}.

Hereafter, we derive the electric thermal conductivity $\kappa$.
According to the linear response theory \cite{Lee,Jujo,Kontani-nu},
\begin{eqnarray}
& \kappa=\displaystyle \frac{-1}{T}\int^\beta_0d\tau\sum_{k_1,k_2}
\frac{1}{i\omega}\langle Q_{k_1}(\tau)Q_{k_2}(0)\rangle 
e^{i\omega \tau}\mid_{\omega\rightarrow0}
 \\
& \hat{Q}_{\mathbf{k}x}(\tau)=\hat{q}_{\mathbf{k}x\uparrow}
+\hat{q}_{\mathbf{k}x\downarrow}+\hat{q}^a_{\mathbf{k}x}
+\hat{q}^{a\dagger}_{\mathbf{k}x}
\end{eqnarray}
where $\hat{Q}_{\mathbf{k}x}$ is the heat current operator
in the superconducting state:
$\hat{q}_{\mathbf{k}x\sigma}$ and $\hat{q}^a_{\mathbf{k}x}$ are given by
\begin{eqnarray}
\hat{q}_{\mathbf{k}x\sigma}=&\omega \hat{v}_{\mathbf{k}x\sigma},
\ \ \
\hat{q}^a_{\mathbf{k}x}=&\omega \hat{v}_{\mathbf{k}x},
\end{eqnarray}
where $\hat{v}_{\mathbf{k}x\sigma}$ represents the ``Fermi velocity''
and $\hat{v}^a_{\mathbf{k}x}$ is the ``gap velocity'' \cite{Lee}:
\begin{eqnarray}
\hat{v}_{\mathbf{k}x\sigma}=&v_{\mathbf{k}x}
c^\dagger_{\mathbf{k}\sigma}c_{\mathbf{k}\sigma},
\ \ \ 
\hat{v}^a_{\mathbf{k}x}=&v_{\mathbf{k}x}^a
c^\dagger_{\mathbf{k}\uparrow}c^\dagger_{-\mathbf{k}\downarrow},
\end{eqnarray}
where $v_{\mathbf{k}x}=v_{\mathbf{k}x}^0+\displaystyle
\frac{\partial \Sigma^n_{\mathbf{k}}}{\partial k_x}$ and 
$v_{\mathbf{k}x}^a=\displaystyle
\frac{\partial \Sigma^a_{\mathbf{k}}}{\partial k_x}$.

As a result, the expression for $\kappa$ in the SC state
with dropping the current vertex correction (CVC)
is given by \cite{Lee,Jujo,Kontani-nu}
\begin{eqnarray}
\kappa&=&\displaystyle\frac{1}{2T}\sum_{\mathbf{k}}\int dz 
\left(-\frac{\partial f(z)}{\partial z}\right)\{2q_{\mathbf{k}x}^2
(G^R_{\mathbf{k}}(z)G^A_k(z) \nonumber \\
&-&F^R_{\mathbf{k}}(z)F^A_{\mathbf{k}}(z)) \nonumber \\
&+&q^{a2}_{\mathbf{k}x}(-G^R_{\mathbf{k}}(-z)G^R_{\mathbf{k}}(z)
-G^A_{\mathbf{k}}(-z)G^A_{\mathbf{k}}(z) \nonumber\\
&+&2F^R_{\mathbf{k}}(z)F^A_{\mathbf{k}}(z)) \nonumber \\
&+&4q_{\mathbf{k}x}q^a_{\mathbf{k}x}(-G^R_{\mathbf{k}}(z)
F^A_{\mathbf{k}}(z)-F^R_{\mathbf{k}}(z)G^A_{\mathbf{k}}(z))\}
 \label{eqn:K}
\end{eqnarray}
where $f(z)=(e^{z/T}+1)^{-1}$.
In the normal state, heat CVC due to Coulomb interaction is small,
as shown in the second-order perturbation theory with respect to $U$
 \cite{Kontani-nu}, and in the FLEX approximation \cite{Kontani-nu-HTSC}.
This fact will also be true below $T_c$ since the particle-nonconserving
four point vertex is much smaller than the particle-conserving one
 \cite{Leggett}.
Therefore, we neglect the heat CVC in the present numerical study.
We find that the first term in eq. (\ref{eqn:K}), 
which is proportional to $\{q_{\mathbf{k}x}\}^2$, is predominant, 
and the other terms which contain the gap velocity, $q_{\mathbf{k}x}^a$, 
are negligibly small.
On the other hand, the charge CVC gives anomalous transport properties
for $R_{\rm H}$ \cite{Kontani-Hall}, $S$ \cite{Kontani-S}, 
magnetoresistance \cite{Kontani-MR-HTSC} and 
Nernst coefficient \cite{Kontani-nu-HTSC} both in hole-doped
and electron-doped systems.
Anomalous transport phenomena due to charge CVC are also observed
in CeMIn$_5$ (M=Co,Rh) \cite{Nakajima2}
and in $\kappa$-(BEDT-TTF)$_2$X \cite{Taniguchi}.

\begin{figure}
\includegraphics[scale=0.45]{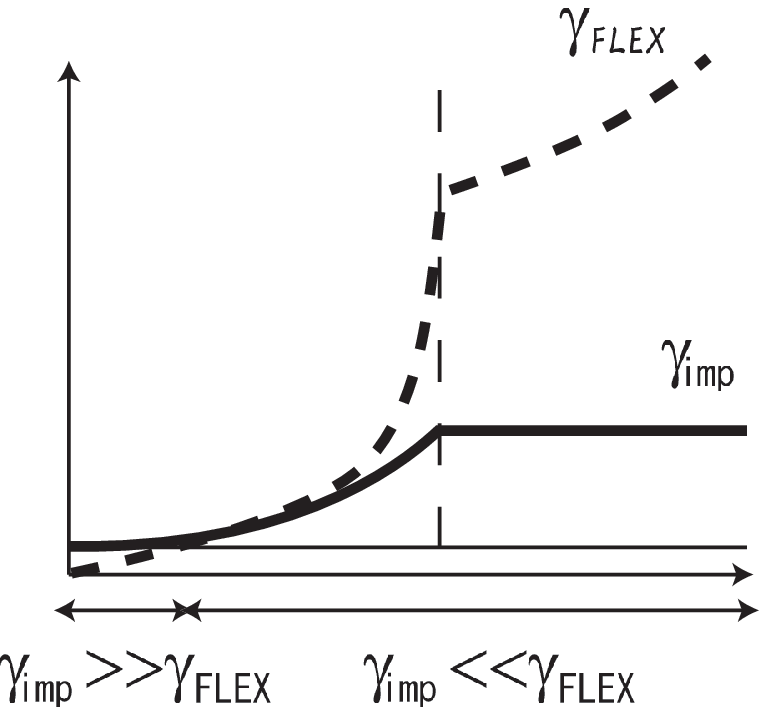}
\caption{Schematic $T$-dependences of $\g_\k^{\rm FLEX}$ 
and $\gamma_{\rm imp}$.
}
\label{fig:schmatic-g}
\end{figure}

Here, we include the QP damping due to impurity scattering $\g_{\rm imp}$
by replacing $\Sigma^{nR}(\mathbf{k},\omega) \rightarrow
\Sigma^{nR}(\mathbf{k},\omega)- i\gamma_{\rm imp}$.
Then, the total QP damping rate is $\g_\k = \g_\k^{\rm FLEX}
+ \gamma_{\rm imp}$ \cite{Hirshfeld,Lofwander}, 
where $\g_\k^{\rm FLEX}= {\rm Im}\Sigma^{nR}(\mathbf{k},\omega)$.
In the $t$-matrix approximation,
$\gamma_{\rm imp}= n_{\rm imp} {\rm Im} \{ -1/(I^{-1}-g_0)\}|_{\w=0}$, 
where $n_{\rm imp}$ is the impurity concentration,
$I$ is the impurity potential, and 
$g_0\equiv\frac1{N}\sum_\k G^R_{\mathbf{k}}$ is the local Green function.
Hereafter, we consider the nearly unitary limit case $I\sim \infty$ where
$\gamma_{\rm imp}(T=0)$ takes a constant value in the self-consistent
calculation \cite{Hirshfeld,Lofwander},
and assume that $\gamma_{\rm imp}\ll \gamma_\k$ at $T=T_{\rm c}$.
In this case, we are allowed to put 
$\gamma_{\rm imp}(T)= \gamma_{\rm imp}(T=0)$ since the $T$-dependence
of $\gamma_{\rm imp}$ affects $\kappa$ near $T_{\rm c}$ only slightly.
A schematic $T$-dependences of $\g_\k^{\rm FLEX}$ and $\gamma_{\rm imp}$
are shown in Fig. \ref{fig:schmatic-g}.

\begin{figure}
\includegraphics[scale=0.38]{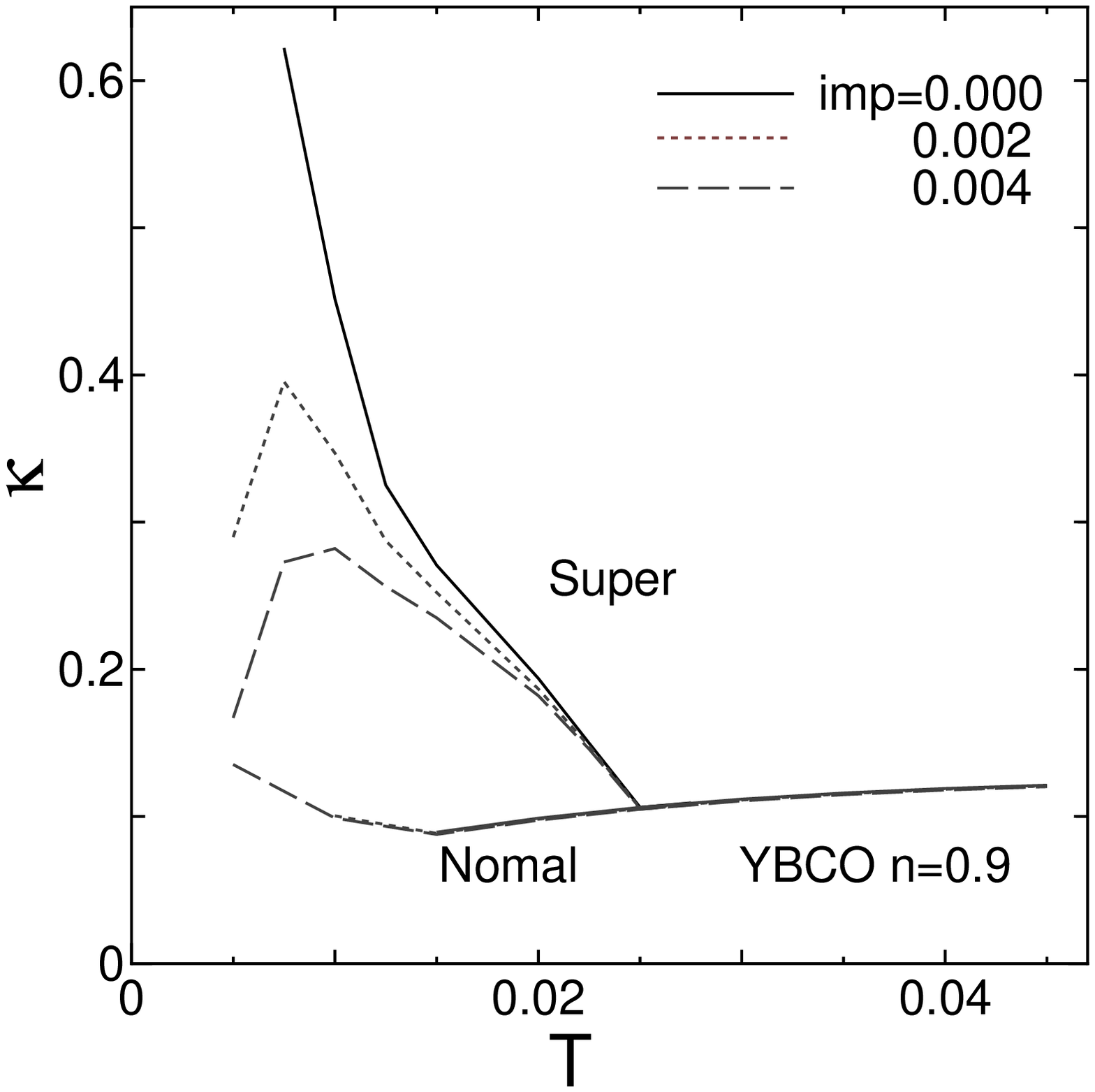}
\includegraphics[scale=0.38]{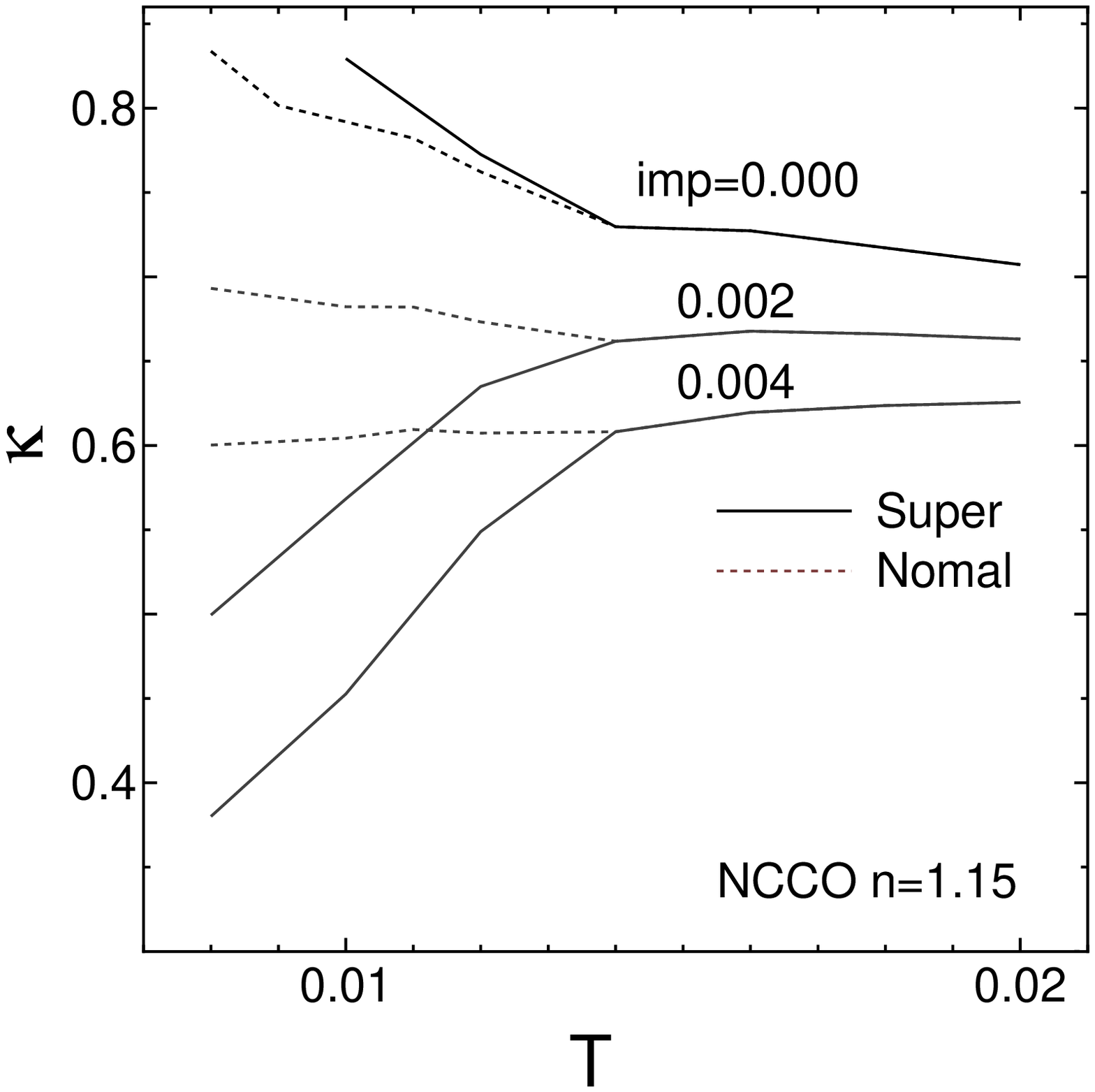}
\caption{Obtained $\kappa$ for YBCO ($n=0.9$; $T_{\rm c}=0.024$) 
and for NCCO ($n=1.15$; $T_{\rm c}=0.014$),
for $\gamma_{\rm imp}=0$, 0.002 and 0.004.
``Normal'' represents $\kappa$ given by the normal state 
FLEX approximation under the constraint $F=\Sigma^a=0$.
}
\label{fig:K}
\end{figure}
Figure \ref{fig:K} represents the temperature-dependence of 
$\kappa$ given by eq. (\ref{eqn:K}).
In YBCO, $\kappa$ increases drastically below $T_{\rm c}$
since the AF fluctuations, which are the origin of inelastic scattering,
are reduced due to the SC gap.
Since $\gamma_\k$ is much larger than $\gamma_{\rm imp}$
at $T>T_{\rm c}$, $\kappa$ in the normal state 
is affected by $\gamma_{\rm imp}$ only slightly.
For $T\ll T_{\rm c}$, on the other hand,
$\kappa$ is suppressed by $\gamma_{\rm imp}$;
$\kappa$ shows the maximum when $\gamma_\k^{\rm FLEX}\sim\g_{\rm imp}$ 
is satisfied at the nodal point.
The obtained result is consistent with experiments \cite{Popoviciu,Ong,Ong2}.
In strong contrast, in NCCO, ``coherence peak'' in $\kappa$ is very 
small even for $\g_{\rm imp}=0$, which is also consistent with experiments 
\cite{Cohn,Fujishiro}.

Here, we discuss the reason why the coherence peak in $\kappa$
is present in YBCO whereas it is absent in NCCO.
Below $T_{\rm c}$, only thermally excited QPs above the SC gap
can contribute to $\kappa$, except at the nodal point.
According to eq. (\ref{eqn:K}), thermal conductivities
in the normal state ($\kappa_n$) and in the line-node SC state 
($\kappa_s$), where $\rho(\e)\propto |\e|$,
are approximately given by
\begin{eqnarray}
\kappa_n \propto& T/\gamma_{\rm cold} 
 \label{eqn:approx-n} \\
\kappa_s \propto& T^2/\gamma_{\rm node}
 \label{eqn:approx-a}
\end{eqnarray}
where $\gamma_{\rm cold}$ and $\gamma_{\rm node}$ represent $\gamma_\k$
at the cold spot above $T_{\rm c}$ and that at the nodal point 
below $T_{\rm c}$, respectively.

\begin{figure}
\includegraphics[scale=0.38]{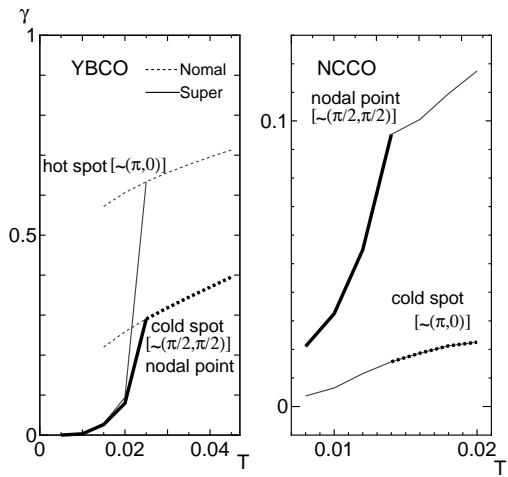}
\caption{$T$-dependence of $\gamma_\k$ for both YBCO and NCCO
for $\gamma_{\rm imp}=0$.
Thick full lines and thick broken lines represent 
$\gamma_{\rm node}$ ($T<T_{\rm c}$) and $\gamma_{\rm cold}$
($T>T_{\rm c}$), respectively.
}
\label{fig:YBCO-NCCO-gamma}
\end{figure}

Figure \ref{fig:YBCO-NCCO-gamma} shows the $T$-dependence of 
$\gamma_\k$ given by the FLEX approximation.
In YBCO, both  $\gamma_{\rm cold}$ and $\gamma_{\rm node}$ are 
given by $\gamma_\k$ at the same point; $\k\approx(\pi/2,\pi/2)$.
As the temperature drops, $\gamma_{\rm cold}$ decreases moderately 
in proportion to $T$ in the normal state.
Below $T_{\rm c}$, $\gamma_{\rm node}$ quickly approaches zero 
since inelastic scattering is suppressed by the SC gap.
As a result, $\kappa$  shows a prominent coherence peak below $T_{\rm c}$,
as recognized by eqs. (\ref{eqn:approx-n}) and (\ref{eqn:approx-a}).
In NCCO, on the other hand, $\gamma_{\rm cold}$ is given by
$\gamma_\k$ at $\k\approx(\pi,0)$, which is different from 
the nodal point of the SC gap; $\k\approx(\pi/2,\pi/2)$.
Since $\gamma_{\rm node}$ is much larger than $\gamma_{\rm cold}$ 
at $T=T_{\rm c}$ in NCCO as shown in Fig. \ref{fig:YBCO-NCCO-gamma},
$\kappa$ is not enhanced in NCCO below $T_{\rm c}$.
Although the numerical accuracy becomes worse for NCCO below $T\sim0.015$,
the obtained result of $\kappa$ will be qualitatively reliable.
In summary, the coherence peak in $\kappa$ is absent even in nodal SC
when the cold spot and the nodal point are different.

\begin{figure}
\includegraphics[scale=0.32]{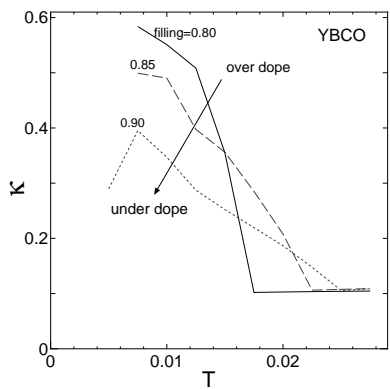}
\caption{$n$-dependence of $\kappa$ for YBCO; $n=0.8\sim0.9$.
}
\label{fig:YBCO-K-n}
\end{figure}

Figure \ref{fig:YBCO-K-n} shows the obtained 
doping dependence of $\kappa$ in YBCO.
In over-doped case ($n=0.8$), the enhancement of $\kappa$
is largest, and it decreases in optimally ($n=0.85$) and under-doped 
($n=0.9$) cases since the cold spot approaches the AFBZ as $n\rightarrow1$.
This tendency is consistent with experiments \cite{Popoviciu}.
Note that in real materials, $T_{\rm c}$ in under-doped case
is smaller than that in optically-doped case.
In the FLEX approximation, however, $T_{\rm c}$ monotonically increases 
as $n\rightarrow1$ since the pseudo-gap state in under-doped region
cannot be described.
The characteristic pseudo-gap phenomena are well described by including the 
self-energy correction due to strong SC fluctuations into the 
FLEX approximation, which is called the FLEX+$T$-matrix approximation
 \cite{Yamada-text,Kontani-nu-HTSC}.

In the present work, we assumed that the inelastic scattering is 
dominant, and neglected the temperature dependence of $\g_{\rm imp}$.
This assumption will be allowed for clean optimally-doped HTSCs.
In dirty samples where elastic scattering is large, we should calculate 
the $T$-dependence of $\g_{\rm imp}$ using the self-consistent
$t$-matrix approximation \cite{Hirshfeld,Lofwander}.
In under-doped systems, however, the $t$-matrix approximation is not 
sufficient since the radius of ``effective impurity potential'' 
is enlarged due to electron-electron correlation, which can be 
described by the $GV^I$-method in Ref. \cite{GVI}.
For a reliable study of $\kappa$ in under-doped systems, 
it will be necessary to take account of residual disorders
using the $GV^I$-method.

In summary, we studied thermal conductivity $\kappa$ in HTSCs.
In the hole-doped case, $\kappa$ shows a prominent ``coherence peak''
below $T_{\rm c}$, whereas it is absent in the electron-doped case.
Based on the FLEX approximation, such a contrasting behavior of $\kappa$ 
is well explained, although both YBCO and NCCO
are pure $d_{x^2\mbox{-}y^2}$-wave superconductors.
We do not have to assume a full-gap state in NCCO 
(such as $d+is$) to explain the absence of a coherence peak,
which originates from the fact that the cold spot (line) in the normal 
state [$\sim(\pi,0)$] is not on the nodal point (line) of the SC gap.
The present study will open the way for the theoretical study of $\kappa$ 
in various interesting unconventional superconductors.

The authors acknowledge fruitful discussions with Y. Matsuda and K. Izawa.
This work was supported by Grant-in-Aid from MEXT.
Numerical calculations were performed at the supercomputer center, ISSP.


\begin{thebibliography}{99}


\bibitem{Sato}  J. Takeda, T. Nishikawa, and M. Sato: 
Physica C {\bf 231} (1994) 293.

\bibitem{Kontani-Hall} H. Kontani, K. Kanki and K. Ueda: 
Phys. Rev. B {\bf 59} (1999) 14723.

\bibitem{Kontani-S} H. Kontani: J. Phys. Soc. Jpn. {\bf 70} (2001) 2840.

\bibitem{Kontani-rev} H. Kontani and K. Yamada: 
J. Phy. Soc. Jpn. {\bf 74} (2005) 155.

\bibitem{izawa} K.~Izawa, H.~Yamaguchi, Y.~Matsuda, H.~Shishido, 
R.~Settai and Y.~Onuki: Phys. Rev. Lett. {\bf 87} (2001) 057002.
\bibitem{Matsuda-rev} Y.~Matsuda, K.~Izawa and I. Vekhter, 
J. Phys.: Condens. Matter {\bf 18} (2006) R705.

\bibitem{Popoviciu} C.P. Popoviciu and J.L. Cohn: 
Phys. Rev. B {\bf 55} (1997) 3155.
\bibitem{Ong} K. Krishana, J. M. Harris, and N. P. Ong:
Phys. Rev. Lett. {\bf 75} (1995) 3529
\bibitem{Ong2} Y. Zhang, N. P. Ong, P. W. Anderson, D. A. Bonn, R. Liang, 
and W. N. Hardy: Phys. Rev. Lett. {\bf 86} (2001) 890.

\bibitem{Movshovich} R. Movshovich, M. Jaime, J.D. Thompson1, C. Petrovic, 
Z. Fisk, P.G. Pagliuso, and J.L. Sarrao: Phys. Rev. Lett. {\bf 86} (2001) 5152.
\bibitem{Kasahara}
Y. Kasahara, Y. Nakajima, K. Izawa, Y. Matsuda, K. Behnia, H. Shishido, 
R. Settai, and Y. Onuki: Phys. Rev. B {\bf 72} (2005) 214515.

\bibitem{Matsuda} Y. Matsuda et al, preprint.

\bibitem{Hirshfeld} P.J. Hirshfeld and W.O. Putikka:
Phys. Rev. Lett. {\bf 77} (1996) 3909. 

\bibitem{Cohn} J. L. Cohn, M.S. Osofsky, J.L. Peng, Z. Y. Li, and R.L. Greene:
Phys. Rev. B {\bf 46} (1992) 12053.
\bibitem{Fujishiro} H. Fujishiro, M. Ikeba, M. Yagi, M. Matsukawa, H. Ogasawara
and K. Noto: Physica B {\bf 219\&220} (1996) 163.


\bibitem{Takahashi} H. Matsui, K. Terashima, T. Sato, T. Takahashi, 
M. Fujita, and K. Yamada: Phys. Rev. Lett. {\bf 95} (2005) 017003.

\bibitem{Hirashima} H. Yoshimura and D.S. Hirashima:
J. Phys. Soc. Jpn {\bf 73} (2004) 2057.

\bibitem{Qazilbash} M.M. Qazilbash, A. Biswas, Y. Dagan, R.A. Ott, 
and R.L. Greene: Phys. Rev. B {\bf 68} (2003) 024502.

\bibitem{Armitage1}
 N.P. Armitage, D.H. Lu, C. Kim, A. Damascelli, K.M. Shen, F. Ronning, 
 D.L. Feng, P. Bogdanov, and Z.-X. Shen:
 Phys. Rev. Lett. {\bf 87} (2001) 147003.
\bibitem{Armitage2}
N. P. Armitage, F. Ronning, D. H. Lu, C. Kim, A. Damascelli, 
K. M. Shen, D. L. Feng, H. Eisaki, Z.-X. Shen, P. K. Mang, N. Kaneko, 
M. Greven, Y. Onose, Y. Taguchi, and Y. Tokura:
Phys. Rev. Lett. {\bf 88}, 257001 (2002).

\bibitem{Lee} A.C. Durst and P.A. Lee: Phys. Rev. B {\bf 62} (2000) 1270.
\bibitem{Jujo} T. Jujo: J. Phy. Soc. Jpn. {\bf 70} (2000) 1349.
\bibitem{Kontani-nu} H. Kontani, J. Phys. Rev. B {\bf 67} (2003) 014408.

\bibitem{Kontani-MR-HTSC} H. Kontani: Phys. Soc. Jpn. {\bf 70} (2001) 1873.
\bibitem{Kontani-nu-HTSC} H. Kontani, Phys. Rev. Lett. {\bf 89} (2002) 237003. 

\bibitem{Leggett} A.J. Leggett: Phys. Rev. {\bf 146} (1965) A1869.

\bibitem{Nakajima2} Y. Nakajima, H. Shishido, H. Nakai, T. Shibauchi, 
K. Behnia, K. Izawa, M. Hedo, Y. Uwatoko, T. Matsumoto, R. Settai, 
Y. Onuki, H. Kontani, and Y. Matsuda: 
J. Phys. Soc. Jpn. {\bf 76} (2007) 027403.

\bibitem{Taniguchi}K.~Katayama, T.~Nagai, H.~Taniguchi, 
K.~Satoh, N.~Tajima and R.~Kato, 
to be published in J. Low Temp. Phys. (2006).

\bibitem{Lofwander} T. Lofwander and M. Fogelstrom: 
Phys. Rev. Lett. {\bf 95} (2005) 107006.

\bibitem{Yamada-text}K. Yamada: {\it Electron Correlation in Metals} 
(Cambridge Univ. Press 2004).

\bibitem{GVI} H. Kontani and M. Ohno: Phys. Rev. B {\bf 74}, 014406 (2006).


\end{thebibliography}
\end{document}